\documentclass{aa}  
\usepackage{graphicx}
\usepackage{txfonts}
\usepackage{natbib}
\bibpunct{(}{)}{;}{a}{}{,} % to follow the A&A style
\graphicspath{{figs/},{./}}

\newcommand{\pdv}[2]{\frac{\partial #1}{\partial #2}}

\newcommand\citen[1]{\citealt{#1}}

\newcommand{\revision}{}
\newcommand{\revisionb}{}
\newcommand{\revisionc}{}
\begin{document} 

\title{Optimization of surface flux transport models for
the solar polar magnetic field}
   \author{K. Petrovay
          \inst{1}
          \and
          M. Talafha\inst{2}%\fnmsep\thanks{Just to show the usage
%          of the elements in the author field}
          }
   \institute{E\"otv\"os Lor\'and University, Department of Astronomy, 
              Budapest, Hungary\\
              \email{k.petrovay@astro.elte.hu}
         \and
              E\"otv\"os Lor\'and University, Department of Astronomy,
	      Budapest, Hungary\\
             \email{m.talafha@astro.elte.hu}
             %\thanks{The university of heaven temporarily does not
              %       accept e-mails}
             }
   \date{Received ; accepted }

% \abstract{}{}{}{}{} 
% 5 {} token are mandatory
 
  \abstract
   {The choice of free parameters in surface flux transport (SFT) models
describing the evolution of the large-scale poloidal magnetic field of
the Sun is critical for the correct reproduction of the polar magnetic
flux built up during a solar cycle, which in turn is known to be a
good predictor of the amplitude of the upcoming cycle. }
   { For an informed
choice of parameters it is important to understand the effect and
interplay of the various parameters and to optimize the models for the
polar magnetic field. }
   {Here we present the results of a large-scale systematic study of the
parameter space in an SFT model where the source term representing the net
effect of tilted flux emergence was chosen to represent a typical, average solar
cycle as described by observations.
}
{Comparing the results with observational constraints on the spatiotemporal
variation of the polar magnetic field, as seen in magnetograms for the last four
solar cycles, we mark allowed and excluded regions in the 3D parameter space
defined by the flow amplitude $u_0$, the magnetic diffusivity $\eta$ and the
decay time scale $\tau$, for three different assumed meridional flow profiles.}
   {Without a significant decay term in the SFT equation (i.e., for
$\tau >10$ yr) the global dipole moment reverses too late in the cycle
for all flow profiles and  parameters, providing independent
supporting evidence for the need of a decay term, even in the case of
identical cycles.  An allowed domain is found to exist for $\tau$
values in the 5--10 yr range for all flow profiles considered.
Generally higher values of $\eta$ (500--800 {\revision km$^2/$s}) are preferred
though some solutions with lower $\eta$ are still allowed.}

   \keywords{Sun -- Dynamo -- Magnetic field -- Solar Cycle}

   \maketitle
%
%________________________________________________________________

\section{Introduction}
\label{sect:intro}

Synoptic maps of the line-of-sight component of the large scale solar
magnetic field have been available on a regular basis since the 1970s.
{\revision Most of this large-scale flux resides in unipolar areas
where one polarity prevails among the magnetic network elements
(\citen{Stix:book}).}
High resolution observations indicate that this field is mostly
concentrated in numerous thin, strong flux tubes that are vertically
aligned 
{\revision (as illustrated e.g. in Fig.~2 of \citen{Orozco2007})}, 
so the true photospheric field is assumed to be radial, and its
amplitude on the solar surface, $B(\lambda,\phi)$ can be derived by a
de-projection of the line-of-sight field as a function of heliographic
latitude $\lambda$ and longitude $\phi$. 
{\revision (A possible smaller contribution to the large-scale field
from a slight polarity imbalance in the small-scale mixed polarity
internetwork fields would not alter this situation, as the orientation
of this field is more isotropic but on average still symmetric around
the radial direction.) }

In order to interpret the evolution of the photospheric field on these
synoptic maps, surface flux transport (SFT) models were developed in
the 1980s 
{\revision (\citen{sheeley1983quantitative})}. 
The models  described the evolution of $B$ by an advective-diffusive
transport equation, advection being ascribed to differential rotation
and poleward meridional flow, and diffusion interpreted as due to the
mixing action of supergranular flows; a source term was included to
represent flux emergence in the form of newly appearing bipolar active
regions (ARs). This early age of SFT model development, reviewed by
\cite{Sheeley:LRSP}, mostly focused on the overall reproduction of the
evolution of the large-scale field during a period of about one solar
cycle (Cycle 21, for which data were available at the time). 
{\revision These models demonstrate that the reversal of the polar
field and of the solar dipole moment  at the middle of activity cycles
and the build-up of a new poloidal field in the late phases of the
cycle originates from the systematic  latitude-dependent tilt of ARs
relative to the azimuthal direction (Joy's law): as this imparts a
meridional component to the AR magnetic field, each AR gives a
contribution to the global solar dipole moment and these contributions
gradually change the Sun's polar magnetic field, after the magnetic
fields of the trailing parts of decayed active regions are transported
to the poles.} 

Interest in SFT models increased again in the 2000s when they were
used for the reconstruction of the long-term variation of the total
solar magnetic flux and open flux, with a view to reconstructing
coronal and interplanetary conditions  {\revision
(\citen{wang2000long}). The long-term solar record clearly indicates
that solar cycles vary in amplitude in a manner that appears to be
random.} As these explorations run SFT models for many solar cycles,
it became clear {\revision (\citen{schrijver2002missing}) that such
variations in cycle amplitude will, for reasons explained
in the previous paragraph,} result in a random walk of the unsigned
amplitude of the solar dipole moment from one cycle to the next, until
it reaches values well exceeding the typical total contribution for
ARs in a cycle, so polar reversal would cease, quenching the dynamo. 
As a workaround, \cite{schrijver2002missing} suggested the
introduction of a sink or decay term of the form $-B/\tau$ in the
transport equation, {\revision tentatively interpreted as an
unobserved process (U-loop emergence).} The decay term was 
{\revision more convincingly} interpreted as a simplistic
phenomenological representation of a 3D effect by
\cite{Baumann+:decayterm} who argued that, owing to the high aspect
ratio of the solar convective zone, the decay time scale of the field
due to vertical diffusion should be two orders of magnitude shorter
than the decay time scale of horizontal diffusion, implying
$\tau=5$--$10$ years. While it has later been suggested that downward
directed pumping in the subsurface layers may suppress vertical
diffusion, this has not been convincingly demonstrated, nor did it
solve the dipole moment drift problem. 
{\revision Intercycle variations in the meridional flow
(\citen{Wang:flowvar}) or in Joy's law (\citen{Cameron+:tiltprecursor})
have also been suggested to solve the dipole moment drift problem,
though the suggested variations are still disputed and the
time periods studied were relatively short.}

A third period of increased activity in SFT modelling started around
2010, with the increased interest in solar cycle prediction following
the abrupt change of the long-term level of solar activity with Cycle
24 {\revision (\citen{pesnell2008predictions})}. As it was realized that
the best physical precursor of the amplitude of an upcoming solar
cycle was the amplitude of the Sun's polar magnetic field at the start
of the cycle (\citen{Petrovay:LRSP}), the next logical step was the
use of SFT models to describe the buildup process of this polar field,
in the hope of extending the temporal range of cycle prediction.  

Any SFT model needs to make assumptions concerning an ill-constrained
function, the effective meridional flow profile $u(\theta)$, as well
as three free parameters: the flow amplitude $u_0$, the diffusivity
$\eta$ and the decay time scale $\tau$. (Other choices, such as
differential rotation or the form of the source, are much better
constrained by observations.) 
{\revision For applications where an exact quantitative reproduction
of the evolving magnetic field is important the appropriate choice of
parameters will be especially crucial.}
The determination of model parameters is a complex issue, to be discussed
in some detail in Section~\ref{sect:motivation} below. We will argue
that a calibration of SFT models to correctly reproduce the
characteristics of the evolution of the polar magnetic field and of
the solar dipole moment is the method that best fits the objectives of
solar cycle prediction. This argument provides the motivation of the
research presented in the rest of the paper. In
Section~\ref{sect:obsdata} we determine the criteria that the SFT
model of a ``typical'' or ``ideal'' solar cycle should satisfy to
qualify as a good representation of the actual evolution of the solar
polar magnetic fields. Then, in Section~\ref{sect:model} we perform a
systematic study on a large grid of SFT models with different
parameter sets and with an idealized but realistic source term
representing a ``typical'' cycle. Comparing the results with the
preset criteria we mark out allowed and excluded domains in the
parameter space. Results of this analysis are discussed in
Section~\ref{sect:discussion}. Section~\ref{sect:concl} concludes the
paper.

%________________________________________________________________

\section{Motivation}
\label{sect:motivation}

Two general approaches have been taken in the choice of parameter
values and flow profiles in SFT models: direct measurements and 
internal calibrations of the model.

Direct measurements of the meridional flow have been made by numerous
researchers with three different methods: Doppler shifts of spectral
lines, {\revision (e.g., \citen{hathaway1996doppler}}), helioseismic
inversions,  {\revision (e.g., \citen{schad2011unified})}, and  local
correlation tracking of moving magnetic flux concentrations
{\revision (e.g., \citen{iida2016tracking})}.  While a general review of
these efforts is far beyond the scope of this paper, we can safely say
that, despite some claims to the contrary, the results can hardly be
considered concordant. There is not even agreement regarding the
latitude range where the overall meridional flow peaks. To cite just a
few recent results: in a time--distance helioseismic analysis
\cite{Chen+Zhao:highpeak} find that the polewards surface flow peaks
at latitudes of $50$--$60^\circ$, whereas \cite{Imada+:midpeak} report
a peak near $45^\circ$ from magnetic feature tracking, while
\cite{Zhao+:lowpeak} find that the flow peaks at a latitude of $\sim
15^\circ$. Part of the reason for the discrepancy is the very
significant cycle dependence of the flow, following a migrating
pattern associated with torsional oscillations
(\citen{Lin+Chou:cycledep.flows}).

{\revision 
Determinations of the surface differential rotation based on
correlation tracking show that small-scale, short-term correlation
tracking leads to results that strongly deviate from the more rigid
rotation of large-scale, long-lived magnetic patterns
(\citen{Wilcox+:patternrot}, \citen{Wang+:patternrot}). This suggests
that small-scale, short-term tracking studies are of limited use for
the determination of the transport properties of large-scale fields. A
study of the effect of transport processes in the deep convective zone
on surface patterns by \cite{Petrovay+Szakaly:2d.pol} also showed that
the evolution of the large-scale surface field is strongly influenced
or even dominated by subsurface processes, rather than by the flows
localized to the surface. Similar conclusions were reached recently by
\cite{Whitbread+:disconn} who find that the apparent diffusion of
surface magnetic fields resulting from AR decay cannot be represented
by the nominal surface value of the diffusivity unless these fields
are somehow disconnected from the magnetic field in the deep
convective zone. All this may well explain why direct determinations
of the magnetic diffusivity generally tend to yield lower values than
internal calibrations of SFT models (see Table 6.2 in
\citen{Schrijver+Zwaan:book}). 
}

In the face of this state of the affairs, internal calibrations of the
SFT model seem to be the preferable way to follow in the determination of parameters. Parameter studies and optimization attempts have been made previously by several authors.  The first more extensive parameter study of the SFT model was performed by
\cite{baumann2004evolution} who considered the effect of varying each parameter, one at a time, in a model without a decay term, with a fixed form of the meridional flow profile and a simplified source term. More recently, a 3x3 model grid in the $(u_0,\eta)$ plane was presented by \cite{Virtanen+:SFT} with a source term taken from
observations and the resulting butterfly diagram was qualitatively
compared to the observed pattern. 

\cite{Lemerle1} and \cite{Whitbread+:SFT} searched for an optimal
combination of parameters (including flow profile) using the PIKAIA
optimization algorithm. These latter studies were the first to apply a
quantitative criterium (the ``merit function'' of the algorithm) to
judge the goodness of the fit between a model result and observations.
The merit used in those studies was essentially the $\chi^2$
difference between observed and simulated magnetic butterfly diagrams.
These two studies differed in the time period considered and in the
way the source term was constructed from observational data, as well
as in some further details, but their results were in general
comparable. Depending on model details, \cite{Whitbread+:SFT} find
that acceptable values of $u_0$, $\eta$ and $\tau$ can fall in the
ranges 7--22 m/s, 220--800 km$^2$/s and 2.5--30 year, respectively. 
At the same time, the characteristics of polar fields were not
particularly faithfully reproduced in the optimized solution. This is
due to the fact that the polar area covers a small fraction of the
solar surface, having little weight in the determination of the merit
applied.

However, as discussed above in the Introduction, the ability to
correctly reproduce the polar magnetic field is critical if SFT models
are to be applied for solar cycle prediction purposes (as opposed to a
general reproduction of observed magnetic butterfly diagrams).
Therefore, in the present work we perform another SFT model parameter
study where the merit is based on a good quantitative reproduction of
the spatiotemporal variation of solar polar magnetic fields.
Furthermore, while PIKAIA is an extremely powerful tool, it offers
little physical insight and yields only limited information on the
extent of the allowed domain in parameter space. Therefore, another
difference relative to previous work is that instead of using an
optimization algorithm or a more limited systematic study, here we
perform a detailed mapping of the 3D parameter space defined by
$(u_0,\eta,\tau)$ for several alternative meridional flow profiles.
This extensive mapping of the parameter space is supported by our
choice to employ a source term representing an idealized cycle (as
discussed below in more detail) instead of constructing the source
based on the observation of a particular solar cycle.

\begin{figure}[hbtp]
\centering
\includegraphics[width=\columnwidth]{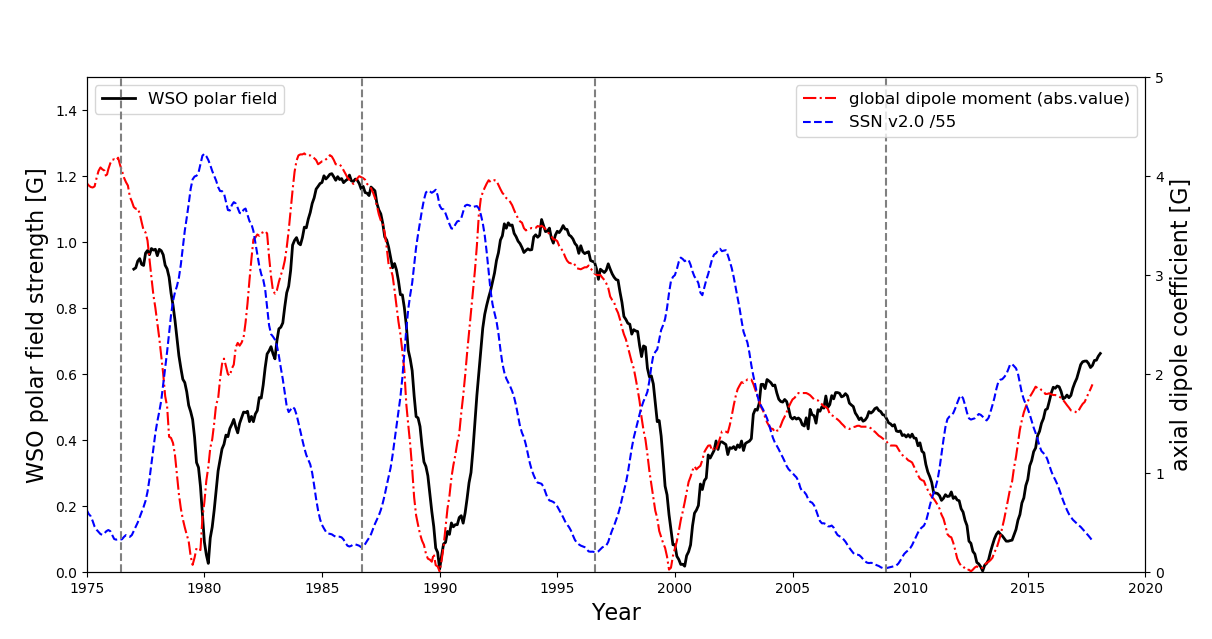}
    \caption{The hemispherically averaged polar field amplitude from
    the WSO data set {\revision (black solid)} and the global axial
    dipole moment {\revision (red dot-dashed)} as a function of time.
    The monthly sunspot number series {\revision (blue dashed)} is
    shown for comparison, with an arbitrary rescaling.  All curves
    were smoothed with a 13-month sliding window. Times of sunspot
    minima are marked by the dashed vertical lines. Global dipole
    amplitudes were obtained by courtesy of Jie Jiang and  represent
    the average of values computed for all available data sets  at the
    given time (\citen{jiang2018predictability}).
}
    \label{fig:polarfieldobs}
\end{figure}

%________________________________________________________________

%_______________________________________________________________
\begin{table*}[hbtp]
   
\caption{Observational constraints on the variation of the poloidal
field in a typical cycle. (Note that $1\sigma$ errors are given here,
but in our parameter mapping $2\sigma$ limits are imposed.)}

\begin{center}
\begin{tabular}{lc}
WSO polar field reversal time counted from sunspot minimum [years]: & $4.33 \pm 0.36$\\
WSO polar field strength at cycle minimum / maximum of WSO polar field
strength: & $0.90 \pm 0.06$\\
Latitude $\lambda_{1/2}$ of the edge of topknot 
[$B(\lambda_{1/2})=B(\theta=0)/2$]: & $70.0^\circ\pm 2.5^\circ$\\
Global dipole moment reversal time counted from sunspot minimum [years]: & $3.44 \pm 0.18$\\
Dipole moment at cycle minimum /  maximum of dipole moment: & $0.84 \pm 0.12$\\
\end{tabular}
\end{center}
\label{table:obsconstraints}

\end{table*}

\section{Observational constraints and choice of merit}
\label{sect:obsdata}

{\revision  The amplitude of the Sun's poloidal magnetic field at
sunspot minimum is known to be a good predictor of the amplitude of
the subsequent solar cycle (\citen{Petrovay:LRSP2}). SFT models offer a
way to to extend the temporal scope of this cycle precursor, provided
they are optimized to best reproduce the observed variation of the
solar poloidal field. In order to determine what are the salient
features of this variation, to be reproduced in the models, in
Figure~\ref{fig:polarfieldobs} we show 
the temporal variation of two measures of the poloidal field,
contrasted with the variation of the sunspot number. In line with
standard practice in solar physics (e.g. in determining sunspot cycle
maxima and minima), all curves have been smoothed with
a 13-month sliding window. The black solid line shows}
the smoothed amplitude of the polar magnetic field
strength measured at Wilcox Solar Observatory (WSO) 
{\revision  averaged over 55--90$^\circ$
latitudes along the central meridian over both poles}. 
The geometry is illustrated in Fig.~1 of \cite{Svalgaard+:polarfield}. 
It is thus approximately given by
\begin{equation}
\tilde{B}_{l}=\frac{1}{1.8\tilde{N}} \int_{\lambda_0}^{\pi/2}  
  B(\lambda,\phi=\phi_{\rm CM},t)  \cos^{2}{\lambda}\, \mathrm{d}\lambda  ,
 \label{eq:WSOB}
\end{equation}
where $\tilde{N}=1-\sin{\lambda_0}$ and  $\lambda_0=55^\circ$ is the
lowest latitude of the polemost pixel. (The factor 1.8 accounts for
magnetograph saturation, cf.\ \citen{Svalgaard+:polarfield}.)
{\revision 
The plotted line is the average of the unsigned amplitude at the two
poles.
}

Also shown is an alternative measure of the amplitude of the poloidal field
component, the {\it axial dipole coefficient}, i.e. the amplitude of the
coefficient of the $Y_1^0$ term in a spherical harmonic expansion of the
distribution of the radial magnetic field strength over the solar disk:
\begin{equation}                       
    D(t) = \frac32 \int_0^{\pi} 
    \overline{B}(\theta,t)\cos\theta\sin\theta\, \mathrm{d}\theta.
 \label{eq:dipmom}
\end{equation}
where $\overline B$ denotes the azimuthally averaged radial magnetic field.
{\revision
In solar physics literature this dipole coefficient is normally referred
to as the solar (axial) dipole moment. (Dipole moment, as usually
defined in physics, has a different dimension but it is indeed
proportional to $D$.)
}

The behaviour of the curves in Figure~\ref{fig:polarfieldobs} shows
that the times of dipole reversal are usually rather sharply defined, 
{\revision 
(in the sense that the dipole moment values below, say,  10\,\% of the peak 
are only observed in time intervals that are much shorter than the cycle 
period).
}
From the plotted data the overall dipole is found to reverse $3.44 \pm
0.18$ years after the minimum, while the WSO field characterizing the
polar contribution to the dipole reverses after $4.33 \pm 0.36$ years.
(The formal $1\sigma$ errors here correspond to the {\revision standard deviations}
calculated for the 4 observed reversals ---admittedly a very  low
number.) The low scatter in these values suggests that the cycle phase
of polar dipole reversal may be a sensitive test of SFT and dynamo
models.
   
In contrast to reversal times, maxima of the dipole amplitude are much
less well defined (occurring $7.27 \pm 1.38$ and $8.33 \pm 1.08$ years
after minimum for the two curves). The curves display broad, slightly
slanting plateaus covering about 3 to 5 years
(\citealt{Iijima+:plateau}); the dipole amplitude at the time of solar
minimum is still typically $84\pm 12\,$\% (global dipole) and $90\pm
6\,$\% (polar fields) of its maximal value, reached years earlier.  

A further well-known empirical result is that  around solar minimum
the poloidal field has a ``topknot'' structure 
{\revision 
(a field distribution that is much more concentrated toward the poles 
than a dipole configuration, \citen{sheeley1989implications}).
\revision Owing to the limited resolution and projection effects, the 
latitude dependence of this field can only be determined indirectly,
exploiting the annual variation of the tilt of the solar axis relative
to the line of sight and comparing the resulting modulation in the
longitudinal field with fits of the form $B\sim\sin^n\lambda$. As
discussed in detail by \citen{Petrie:LRSP}, various determinations of
$n$ have yielded values in the range 8--9. As in our models the
field profiles are not necessarily well fitted by $\sin^n\lambda$ we
characterize the extent of the topknot with the latitude $\lambda_{1/2}$ 
where $B$ drops  to half its value at the pole. The observational
constraints indicating that $n>7$ imply $\lambda_{1/2}>65^\circ$.
The upper limit may be less well defined due to the limited
resolution; so here we adopt $\lambda_{1/2}<75^\circ$ --- a rather
generous value compared to the $69^\circ$ that corresponds to $n=10$.
}

These empirical limits on polar field variation are summarized in 
Table~\ref{table:obsconstraints},
For use in solar cycle prediction
based on the polar precursor method we require that SFT models of a
``typical'' solar cycle must satisfy the listed 5 constraints. In view of
the rather high uncertainty in the value of the mean related to the
very low number of observed cycles, in our parameter study we will we
extend the allowed domain to $\pm 2\sigma$. (Note that, as $\sigma$ is
the {\revision standard deviation of the data}, the uncertainty of the 
mean is only half that -so our study actually allows for a generous
$\pm 4\sigma$ uncertainty in the value of the mean.)

{\revision 
We reiterate that these values were determined for the last four solar
cycles (21--24), Figure~\ref{fig:polarfieldobs}, after smoothing the
data with a 13-month sliding window, and the errors given in 
Table~\ref{table:obsconstraints} are standard deviations calculated
for the four cycles 21--24. E.g., the reversal times of the WSO field
amplitudes for the individual cycles, counted from the last minimum, 
are  4.50, 3.92, 4.08, and 4.83 years, while for the dipole moment
values they were  3.42, 3.27, 3.33, and 3.75 years, respectively. 
For $\lambda_{1/2}$ the nominal values shown in the table correspond
to a fiducial $\pm 2\sigma$ interval of $65^\circ$--$75^\circ$,
adopted as discussed above.
}

\begin{figure}[htbp]
\centering
\includegraphics[width=\hsize]{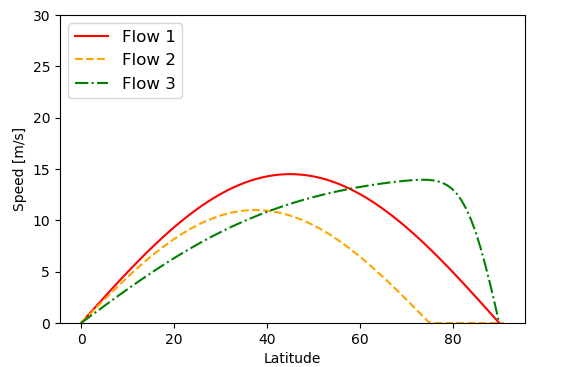}
    \caption{% 
Meridional flow profiles used in the paper. } 
\label{fig:circul}
\end{figure}

\section{Model}
\label{sect:model}

As our intention here is to model ``typical'' or ``average'' solar
cycles, our source function will not consist of individual ARs but a
smooth distribution representing the probability distribution of the
emergence of preceding ($p$) and following ($f$) polarities on the
solar surface. By its nature this source is then axially symmetric, so
our whole SFT model will be reduced to one dimension:
\begin{eqnarray}
\label{eq33}
 \pdv Bt &=& 
 \frac{1}{R\cos{\lambda}}\pdv{}\lambda\left[Bu(\lambda)\cos{\lambda}\right]
 \nonumber \\
 &&+\frac{\eta}{R^2\cos{\lambda}}
 \pdv{}\lambda\left(\cos{\lambda}\pdv B\lambda\right)
 -\frac{B}{\tau} + S(\lambda,t)
\label{eq:transp}
\end{eqnarray}
where $R$ is the solar radius and $S$ is the source representing flux emergence.

\subsection{Meridional flow}

For the meridional flow we consider three different profiles that have been used
in SFT or dynamo models for cycle prediction (Fig.~\ref{fig:circul}).

{\it Flow 1:} a simple sinusoidal profile
\begin{equation}
\label{eq4}
    u_{c}=u_{0}\sin(2\lambda  )
\end{equation}
This profile was used e.g. by \cite{dikpati2006predicting} and
\cite{cameron2007solar}, with $u_0=14.5/,$m$/$s, $\eta=300\,$km$^2/$s and
$\tau=5.6$ yr.

{\it Flow 2:} a sinusoidal profile with a dead zone around the poles,
\begin{equation}
\label{eq5}
    u_{c} =
\left\{
 \begin{array}{ll}
 u_{0}\sin(\pi\lambda/\lambda_{0})  
     & \mbox{if } |\lambda| < \lambda_{0} \\
 0 & \mbox{otherwise } 
 \end{array}
\right.  
\end{equation}
This profile was used by \citep{van1998magnetic} and in numerous papers by  the
MPS/Beijing group (\citep{Cameron+:tiltprecursor}, \citep{jiang2014magnetic},
\cite{Jiang+Cao} ), with $\lambda_0=75^\circ$, 
$u_0=11/,$m$/$s, $\eta=250\,$km$^2/$s and $\tau=\infty$.

{\it Flow 3:} a profile peaking at high latitudes, \begin{equation}
\label{eq6} u_{c}(R,\theta) = \frac{u_{0}}{u_0^\ast}
\mbox{erf}\,(V\cos{\lambda})\, \mbox{erf}\,(\sin{\lambda}) \qquad V=7 
\end{equation} This profile was used by \cite{Lemerle2}.
$u_0^\ast=0.82$ is a normalization factor defined so that $u_{0}$
gives the maximum meridional flow velocity, like in the other
profiles.

\begin{figure*}[htbp]
\centering
\includegraphics[width=\hsize]{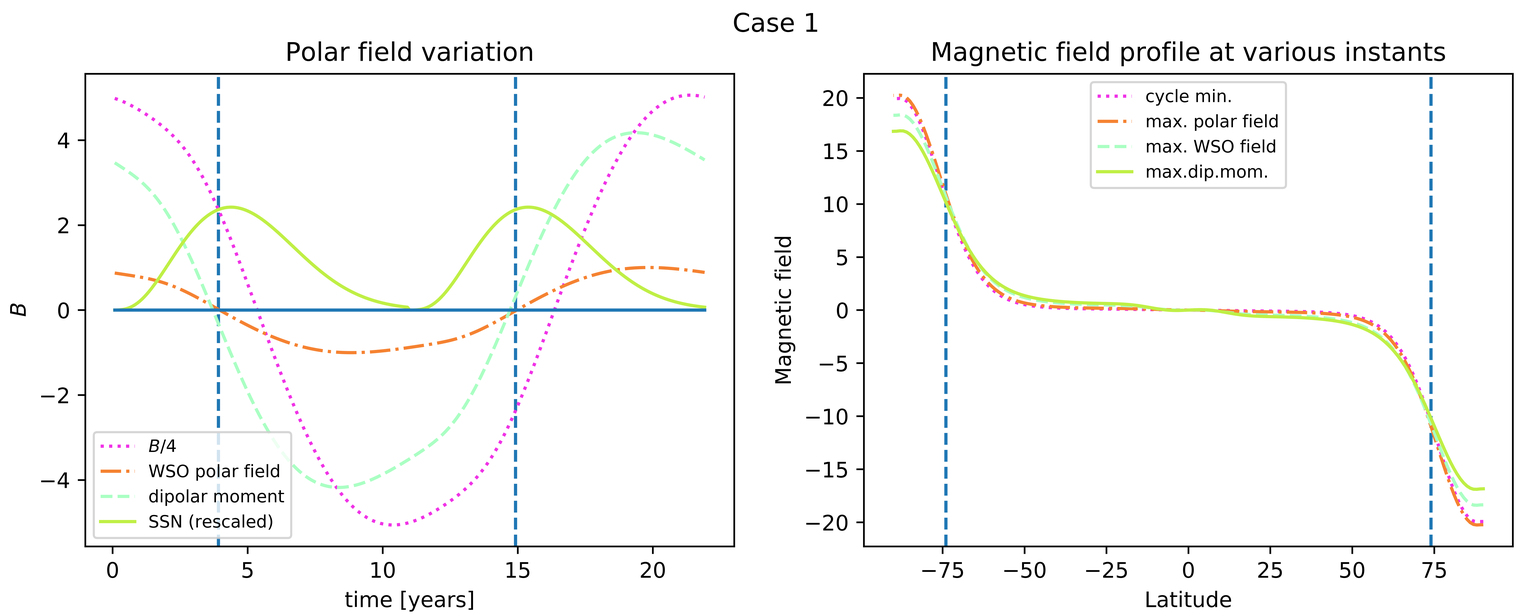}
    \caption{% 
Example of a solution satisfying all observational constraints on the polar
magnetic field: temporal variation and latitudinal profiles
{\revisionb for $u_0=10\,$m$/$s, $\eta=600\,$km$^2/$s and $\tau=7\,$yr
{\revisionc (flow 1)}. 
Time variations of various quantities are shown in the left panel:
in the label of the dotted curve $B$ refers to the field amplitude at
the pole; vertical dashed lines mark the reversals of the WSO polar
field. 
The right panel shows latitudinal profile of the magnetic field at various 
instants as indicated; vertical dashed lines mark the latitudes where
$B$ equals half its value at the poles at cycle minimum.}} 
\label{fig:goodplot}
\end{figure*}

\begin{figure*}[htbp]
\centering
\includegraphics[width=\hsize]{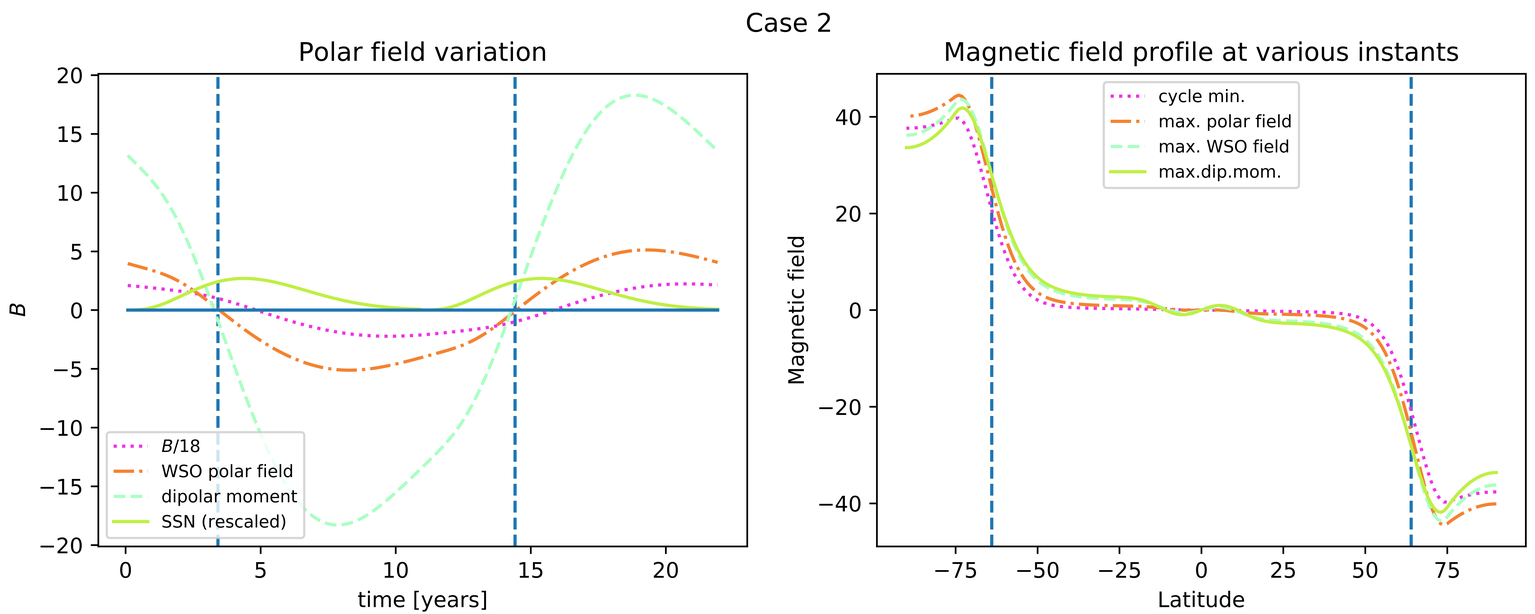}
    \caption{% 
Example of a solution at odds with several observational constraints
on the polar magnetic field: temporal variation and latitudinal
profiles. 
{\revisionb for $u_0=12.5\,$m$/$s, $\eta=500\,$km$^2/$s and
$\tau=5\,$yr {\revisionc (flow 2)}. 
Notations as in Fig.~\ref{fig:goodplot}.}}
\label{fig:badplot}
\end{figure*}

\begin{figure*}[htbp]
\centering
\begin{tabular}{cc}
\includegraphics[width=0.5\hsize]{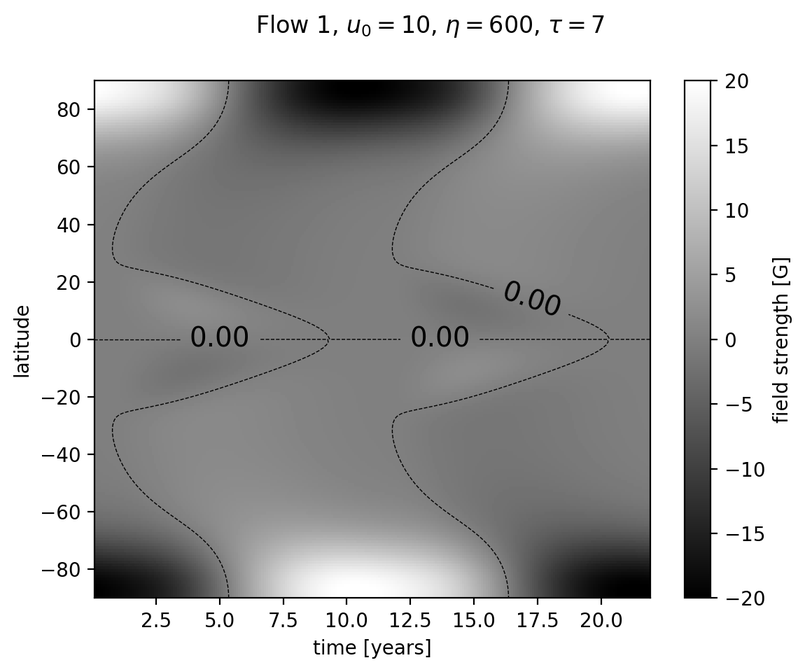} & 
\includegraphics[width=0.5\hsize]{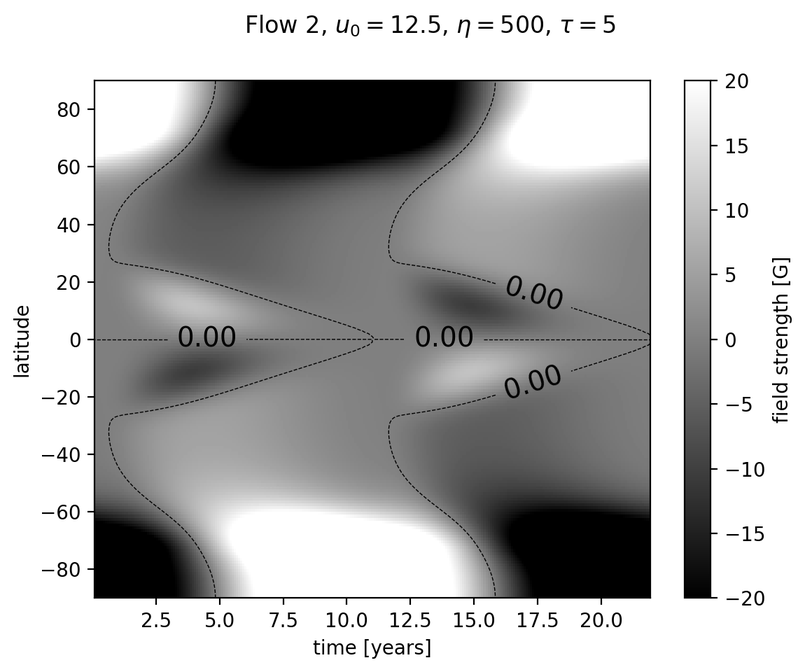}
\end{tabular}
    \caption{% 
Butterfly diagrams {\revision (plots of the field strength against time 
and latitude)} of the example solutions shown in figures
\ref{fig:goodplot} and \ref{fig:badplot}.} 
\label{fig:goodbadbfly}
\end{figure*}

\subsection{Source term}
\label{sect:source}

Our source term is a smooth distribution representing the probability
distribution of the emergence of $p$- and $f$- polarities on the solar
surface. Similar to the approach of \cite{dikpati2006predicting} and
\cite{cameron2007solar}, we represent the source by a pair of rings of
opposite magnetic polarity:
\begin{eqnarray}
S(\lambda,t)&=&
kA_m S_1(t)
S_2\left[\lambda;\lambda_0(t)-\Delta\lambda(t),\delta\lambda\right]
\nonumber \\
&&-kA_m S_1(t)
S_2\left[\lambda;\lambda_0(t)+\Delta\lambda(t),\delta\lambda\right] 
\nonumber \\
&& +kA_m S_1(t)
S_2\left[\lambda;-\lambda_0(t)-\Delta\lambda(t),\delta\lambda\right] 
\nonumber \\
&& -kA_m S_1(t)
S_2\left[\lambda;-\lambda_0(t)+\Delta\lambda(t),\delta\lambda\right]
\end{eqnarray}
where $k=\pm 1$ is a factor alternating between even and odd cycles.
$A_m$ is an arbitrary amplitude depending on the flow profile (0.003,
0.015 and 0.0005 for our three profiles, respectively) used to ensure
that the resulting polar field amplitude roughly agrees with
observations (even though the amplitudes produced in this linear model
are not used as a merit in the optimization).  The latitudinal profile
$S_2\left[\lambda;\lambda_0(t),\delta\lambda\right]$ is a Gaussian
with a fixed full width at half-maximum of $2\delta\lambda=6^\circ$,
migrating equatorward during the course of a cycle:
\begin{equation}
S_2(\lambda;\lambda_0,\delta\lambda) = 
\exp\left[-(\lambda-\lambda_0)^2/2\delta\lambda^2\right]
\end{equation}

The latitudinal separation of the rings is a consequence of Joy's law:
\begin{equation}
2\Delta\lambda=0.5\frac{\sin\lambda}{\sin 20^\circ}
\end{equation}
while their trajectory during the course of a cycle is given by a quadratic fit
derived by \cite{Jiang+:1700a} from many observed solar cycles:
\begin{equation}
\lambda_0 [{}^\circ] = 26.4 - 34.2 (t/P) + 16.1(t/P)^2
\label{eq:quadratic}
\end{equation}
where $P=11\,$year is the cycle period.

Finally, the time profile of solar activity in a typical cycle was determined
by \cite{hathaway1994shape} from the average of many cycles as
\begin{equation}
    S_1(t)= at^{3}_{c}/[\exp(t^{2}_{c}/b^{2})- c]
\end{equation}
with $a = 0.00185$, $b = 48.7$, $c= 0.71$, where $t_c$ is the time since the
last cycle minimum.

In addition to the quadratic source term described above, an alternate form of
the source term was also considered, to check the sensitivity of the results on
this choice. The alternate form had a linear trajectory instead of equation
(\ref{eq:quadratic}) and a constant latitudinal separation of $1^\circ$ between
the flux rings, while in other respects it was identical to the above.

\subsection{Model grid}

For each choice of meridional flow profile and source function we
computed 1105 models with different parameter combinations. $u_0$ was
allowed to vary between 5 and 20 m/s in steps of 2.5; $\eta$ varied
from 50 to 750 km$^2/$s in steps of 50, while $\tau$ varied from 2 to
10 years in steps of 1 year, and two additional values ($\tau=20$ and
$\tau=100$) were also considered, the latter value being effectively
equivalent to a negligible decay term in equation (\ref{eq:transp}).
The transport equation was solved with a simple explicit code, making
sure that the resolution ensures flux conservation to an acceptable
degree not to appreciably influence the results. A latitudinal grid
spacing of 0.5 degree and a timestep of 6 hours was found to suffice
in most cases. Our code is made available on
GitHub\footnote{\url{https://github.com/kpetrovay/Polar-SFT}} to
ensure reproducibility of the results. Starting from an arbitrary
(dipole) initial state, the calculations were run for 20 solar cycles,
until very nearly steady oscillations set in. At that point we
evaluated whether a given model complies with the imposed
observational constraints listed in Table\ref{table:obsconstraints}.

\subsection{Parameter maps}
\label{sect:parammaps}

Figures~\ref{fig:goodplot}, \ref{fig:badplot} and
\ref{fig:goodbadbfly}  show examples of a ``good'' and a ``bad''
solution. 
{\revision
The left-hand panels of Figures~\ref{fig:goodplot} and
\ref{fig:badplot} present the time variation of the actual value of
the polar magnetic field strength $B(\theta=0)$, the value measured by WSO
[eq.~(\ref{eq:WSOB})] and the global dipole moment
[eq.~(\ref{eq:dipmom})]. The right-hand panels in these figures, in
turn, display the latitudinal profiles of $B$ at solar
minimum and at the instants when the measures of polar field amplitude
plotted in the left panels take their maxima.
Figure~\ref{fig:goodbadbfly} displays the spatiotemporal variation of
$B$ against time and latitude, i.e. the magnetic butterfly diagram
characterizing these simulations.
}

While at first sight similar, these plots indicate that in
the second case the polar topknot is too broad and the polar reversal
occurs too early in comparison with the observational constraints
listed in section ~\ref{sect:obsdata}, imposed with a generous
limit of $2\sigma$ (or $4\sigma$ in terms of the uncertainty of the
mean, cf.\ discussion in Section~\ref{sect:obsdata}).

{\revision 

For the two cases plotted, $\lambda_{1/2}$  takes the value $71^\circ$
and $64^\circ$, respectively, while the WSO field is reversed $4.0$
and $3.4$ years after the minimum. Comparing this  with the values in
Table~\ref{table:obsconstraints}, both values fall within the allowed
$\pm 2\sigma$ interval in the first case, while both fall outside it
in the second. The second set of parameter is therefore discarded as
incompatible with the observational constraints on the polar field,
while the first model is admissible by all five constraints.

\begin{figure*}[tbp]
\centering
\includegraphics[width=\hsize]{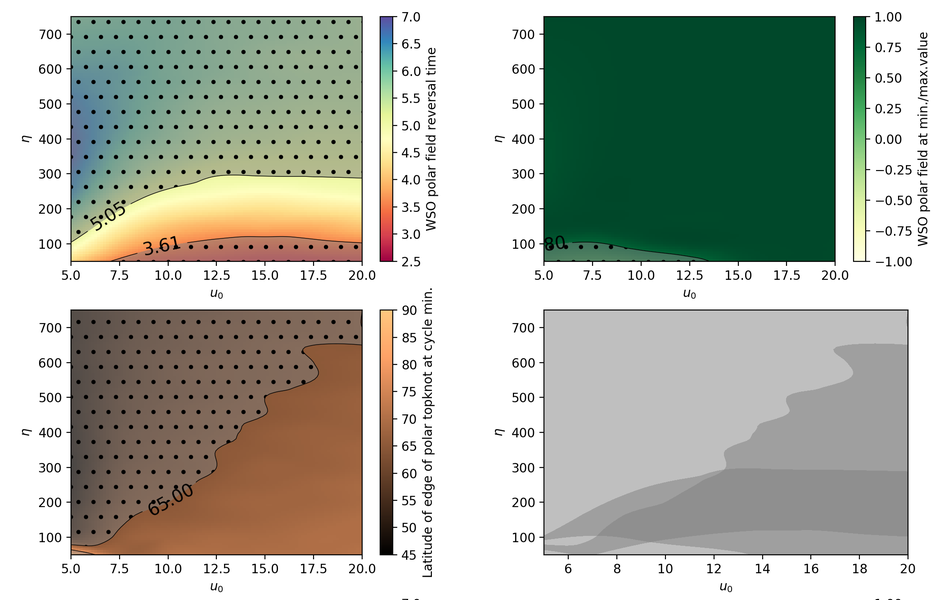}
    \caption{% 
Set of maps of the $u_0$--$\eta$ parameter plane for the case of
$\tau=100$ yr, flow type 2. Colour coded maps show the distribution of
three merit variables, with the excluded domain masked out (less vivid
colours {\revisionb and dotted}). Allowed domains  are shown combined 
in grey in the middle right panel. Units are m/s for $u_0$, km$^2$/s
for $\eta$, years for time and degrees for latitude.} 
\label{fig:maps_case2_100}
\end{figure*}

\begin{figure*}[tbp]
\centering
\includegraphics[width=\hsize]{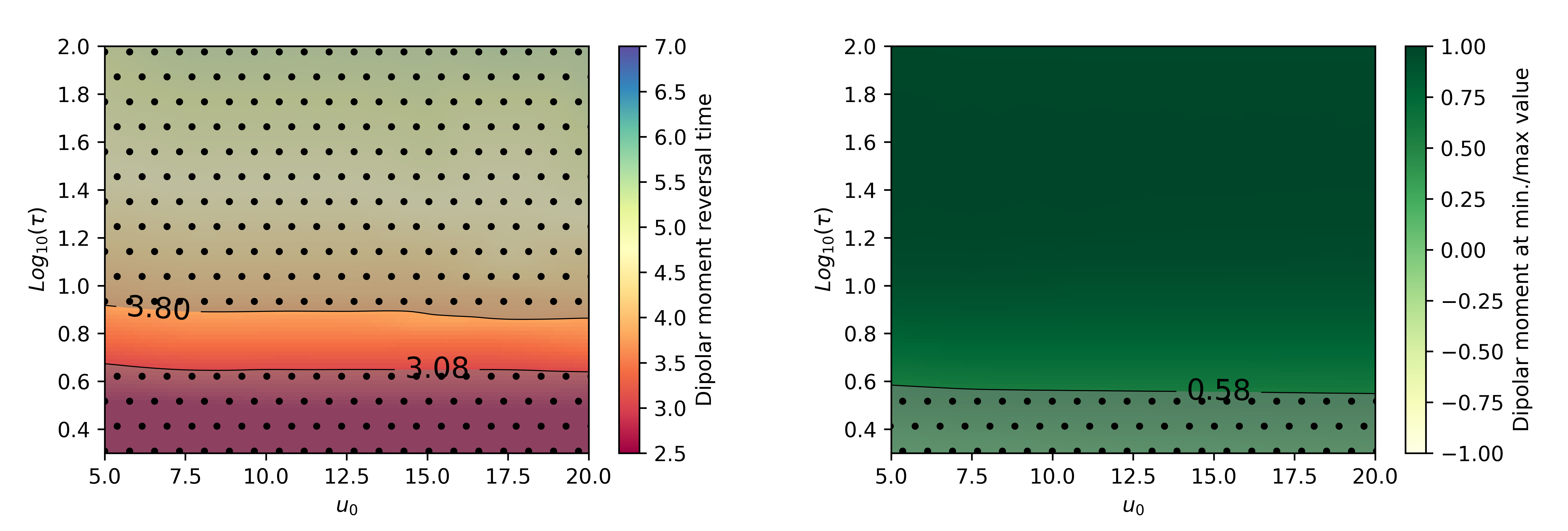}
\includegraphics[width=\hsize]{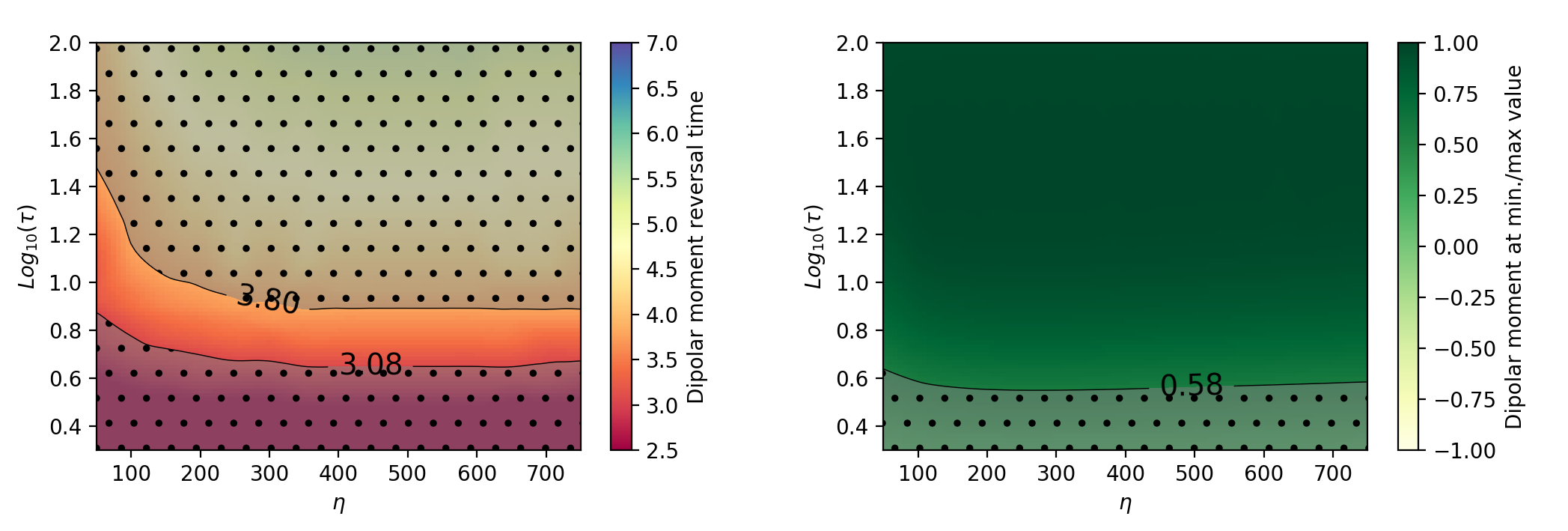}
    \caption{% 
Set of maps of the $u_0$--$\tau$ parameter plane for the case of
$\eta=500$ (top), and of the $\eta$--$\tau$ plane for $u_0=10$,
showing allowed and excluded domains for the dipole reversal time
(left) and ratio of dipole moment at cycle minimum to the maximal
dipole moment (right). Units are m/s
for $u_0$, km$^2$/s for $\eta$ and years for $\tau$.
} 
\label{fig:maps_tauvar}
\end{figure*}

Such comparisons can be performed for each parameter combination in
our model grids, and as a result, admitted and discarded domains in
the 3D ($u_0, \eta, \tau$) parameter space can be identified. In order
to visualize these domains, we construct 2D maps of slices of the
parameter space with fixed values of $\tau$. In these 2D maps the
allowed  domain is shown unmasked, in more vivid colors, while the
excluded domain is shown through a semitransparent {\revisionb dotted}
grey mask, in pale colors. One set of such maps for flow profile 2 at
$\tau=100$ yr is shown in Figure~\ref{fig:maps_case2_100}, where the
colored panels represent the first three merit parameters listen in
Table~\ref{table:obsconstraints} on the $u_0$--$\eta$ plane. The
allowed domains are shown combined in the middle right panel, the
darkest gray showing the domain allowed by all three criteria. 
}

{\revision 
The last two merit criteria related to the global dipole moment are
not included in this plot partly for clarity and partly because these
criteria were found to be most sensitive to $\tau$, so in slices taken
at fixed values of $\tau$ either the allowed or the excluded domain
tends to cover the whole plane. This is separately borne out in
Figure~\ref{fig:maps_tauvar} where slices of the 3D parameter space
orthogonal to the slices in Figure~\ref{fig:maps_case2_100} are shown,
mapping the distribution of the merit parameters associated with the
dipole moment. It is apparent that these parameters are indeed mainly
controlled by $\tau$: for shorter values of $\tau$ the global field
decays faster, so it reverses earlier and declines more significantly
from its maximum to the end of the cycle.
}

%________________________________________________________________

\section{Discussion}
\label{sect:discussion}

One immediate conclusion from these studies follows directly from our above
mentioned observation that the time variation of the dipole moment primarily
depends on $\tau$. In fact we find that for higher values of $\tau$, i.e. {\it
without a significant decay term in the transport equation, the global dipole
moment invariably reverses too late in the cycle.} This is seen in
Fig.~\ref{fig:maps_tauvar}; the equivalent plots look similar also for
other flow types and other values of $u_0$ or $\eta$.

In contrast to this, we find that a range of shorter decay times exists where,
with our realistic source function, all five merit criteria can be satisfied in
some part of the parameter space. This range of $\tau$ lies somewhere between 5
and 10 years for all three flow types considered (though exact limits of the
range depend on the flow profile).

All this can be taken as evidence for the need of a decay term in the transport
equation for the realistic modelling of the time variation of the global dipole
moment in an average cycle. As in our models intercycle variations are not
present, this line of evidence is independent from the drift in dipole
moment that originally motivated the introduction of the decay term
{\revision 
(\citen{schrijver2002missing} ---cf. discussion in Section~1 above).
}

\begin{figure*}[tbp]
\centering
\includegraphics[width=\hsize]{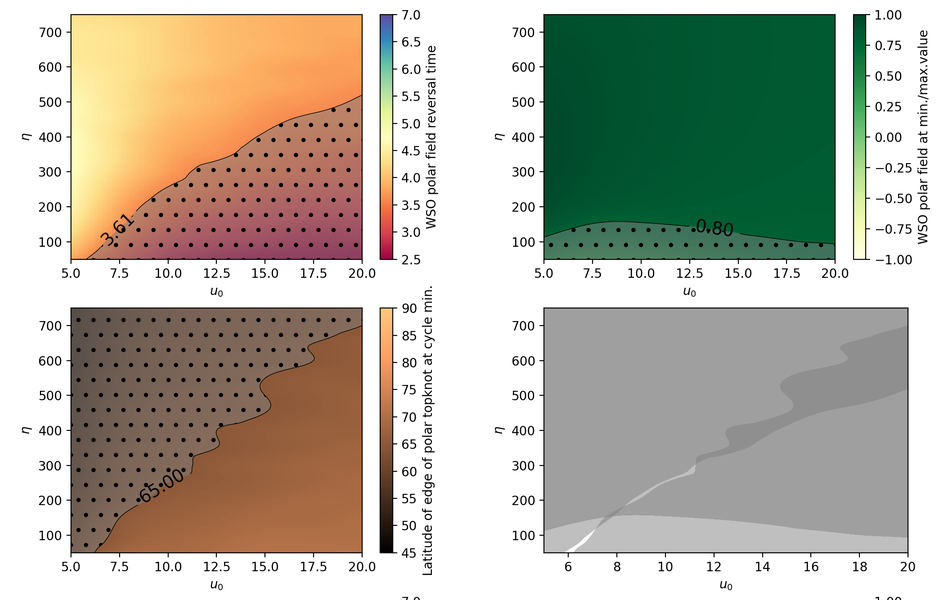}
    \caption{% 
Same as Fig.~\ref{fig:maps_case2_100} for the case of $\tau=7$,
flow type 2.
}
\label{fig:maps_case2_7}
\end{figure*}
 
{\revision
For the allowed solution with $\tau$ in the range $5$--$10$ years, 
higher diffusivity values (500\,km$^2$/s and above) are generally
favoured, i,e. the bulk of the allowed domain lies here.
In the case of the flow with a polar dead zone (flow type 2, 
Fig.~\ref{fig:maps_case2_7}) the value
of $u_0$ generally also needs to be high ($\ga 15\,$m/s) or the topknot 
becomes too broad; however, a narrow band of allowed solutions along
a line of constant $\eta/u_0$ ratio also exists.
}

From this, the parameter combination $u_0=11\,$m/s and
$\eta=250\,$km$^2/$s used by \cite{Cameron+:tiltprecursor} or
\cite{Jiang+Cao} stands out as admissible with a $\tau$
value of about 8 years, while for $\tau\ga 10$ it produces too late
dipole reversals in the ``average'' cycle studied here.

\begin{figure*}[tb]
\centering
\includegraphics[width=0.95\hsize]{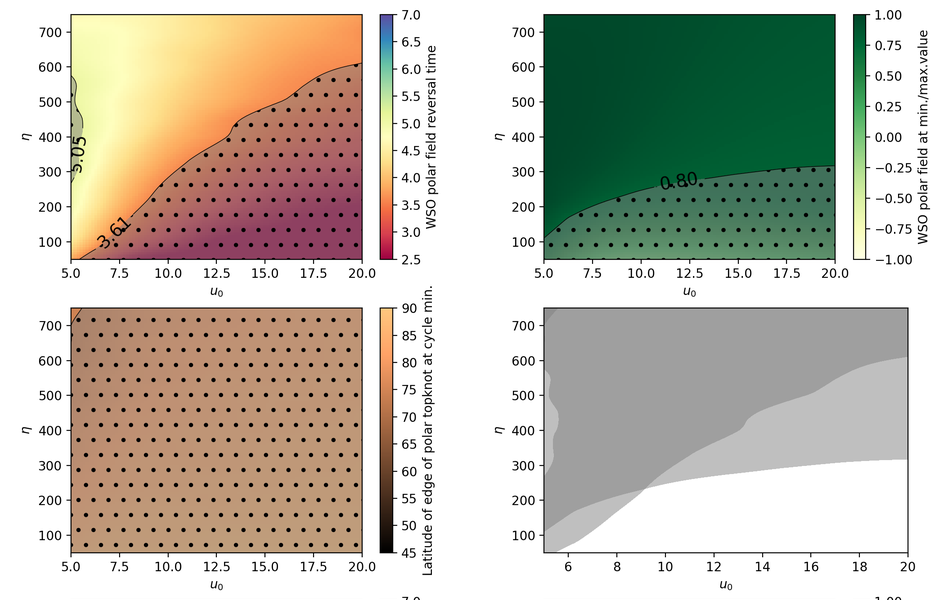}
    \caption{% 
Same as Fig.~\ref{fig:maps_case2_100} for the case of $\tau=10$,
flow type 3.} 
\label{fig:maps_case3_10}
\end{figure*}

\begin{figure*}[bt]
\centering
\includegraphics[width=0.95\hsize]{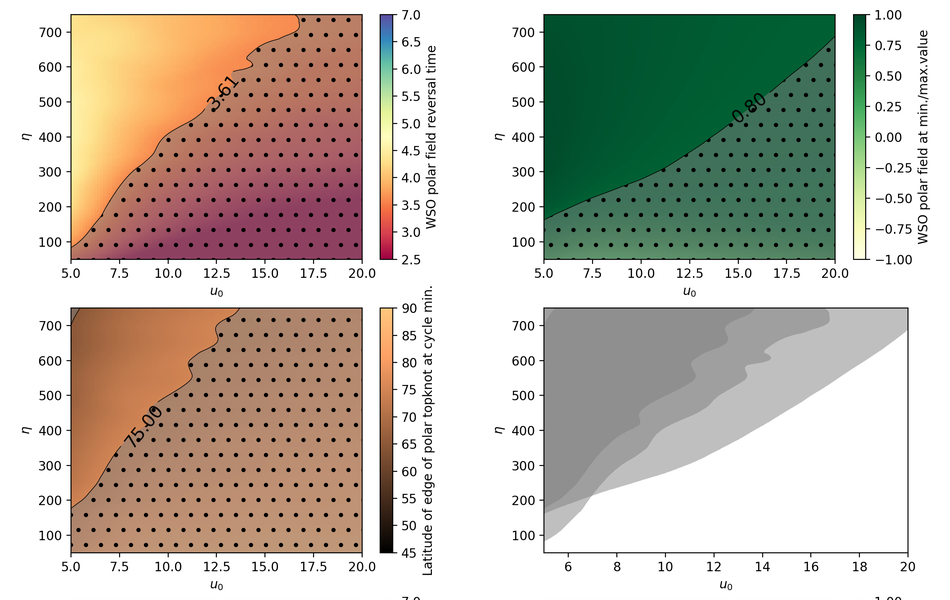}
    \caption{% 
Same as Fig.~\ref{fig:maps_case2_100} for the case of $\tau=7$,
flow type 1.} 
\label{fig:maps_case1_7}
\end{figure*}

Turning our attention now to the pole-reaching flows (type 1 and 3), we find
that the allowed domain that exists in the range $\tau\simeq 5$ to $9$
invariably lies in the top left part of our parameters plane, i.e. combinations
with higher diffusivity and lower flow speed are preferred. This is mainly due
to the topknot width constraint: it was already shown by
\cite{Sheeley+:topknot} that the topknot width for pole-reaching flows scales
with $\eta/u_0$, thus becoming too broad for higher values of this parameter.

In particular, with the flow profile obtained in the  2$\times$2D
dynamo model of \cite{Lemerle2} polar fields will only show fully
solar-like behaviour for $\eta/u_0\ga 150$ (e.g. for $u_0\ga 5\,$m/s
$\eta\ga 700\,$km$^2/$s) and $\tau\sim 5$--$9$ yr. In contrast, the
values {\revisionb $u_0= 17\,$km$/$s, $\eta=600\,$km$^2/$s and $\tau=
10\,$yr,}, 
resulting from their globally optimized
solution, yield a too narrow topknot and a delayed dipole reversal
(Fig.~\ref{fig:maps_case3_10}). The reason why this parameter set was
obtained by \cite{Lemerle2} as an optimalized solution clearly lies in
their different choice of merit and their combined optimization of an
SFT model and a coupled dynamo model, instead of using a source term
modeled on observations.

Overall, the admitted domain is found to be the most extensive for a simple
sinusoidal flow profile with $\tau\simeq 7$ (Fig.~\ref{fig:maps_case1_7}): in
this case, practically the whole upper left half of the $u_0$--$\eta$ plane is
allowed. Taken by themselves, then, the polar field constraints considered in
this paper would suggest the use of this simple flow model for cycle
prediction purposes with a $\eta/u_0$ ratio exceeding about 50.

As mentioned at the end of subsection \ref{sect:source}, for
comparison we also considered a less realistic choice of the source
function, with a linear decrease in AR latitudes during the cycle and
no accompanying variation in the tilt. Results for this case are
available from the GitHub repository. It was found that there exist no
parameter combinations where all five merit criteria can be
simultaneously satisfied in this case. This shows that satisfying our
criteria is non-trivial, and the very existence of admissible
solutions indirectly supports the correctness of our choice of the
source function.

%________________________________________________________________

\section{Conclusion}
\label{sect:concl}

In this paper we mapped the parameter space of surface flux transport
models from the point of view of the spatiotemporal variation of the
polar field resulting with a source term representing an average solar
cycle, marking the allowed domain compatible with observational
constraints. In Section~\ref{sect:motivation} we argued that other
methods of constraining parameter space are less well suited for
applications of SFT to solar cycle prediction. 

A key finding from this work was that without a significant {\revision decay}
term in the SFT equation (i.e., for $\tau >10$ yr) the global dipole
moment reverses too late in the cycle for all flow profiles and 
parameters. This provides independent supporting evidence for the need
of the inclusion of a decay term in SFT models even for identical
cycles.

The question may arise whether this conclusion is related to our use
of a rather restricted range (about $\pm 4$ months) allowed for the
dipolar moment reversal time, derived from the average of all
available observational datasets. Indeed, considering individual
datasets, the dipole reversal often deviates by several months (in
rare cases by even a year) from the averaged set. However, our
parameter maps 
{\revisionb for the dipole moment reversal time (e.g., 
Fig.~\ref{fig:maps_tauvar}) show that for $\tau=100$}
no reasonable extension of the allowed range can account for
the delay in the dipole reversal time which in these cases well
exceeds two years in all cases. We further note that anecdotal
evidence exists in the literature for cases where the dipole reversal
time of one individual cycle was apparently well reproduced in SFT
models without a decay term {\revision (\citen{devore1985numerical}, 
\citen{wang1989evolution}, \citen{Jiang+Cao})}. While none of these
studies  was comprehensive enough to draw general conclusions beyond
the particularities of a given cycle, the roots of this apparent
puzzle clearly need to be understood. It is worth noting that recently
\cite{Virtanen+:SFT} also confirmed the need for a decay term in
order  to attain polar reversal in Cycle 24.

Turning now to models with a decay term, an allowed domain was found
to exist for $\tau$ values in the 5--10 yr range for all flow profiles
considered. Generally higher values of $\eta$ (500--800) are
preferred, though some solutions with lower $\eta$ are still allowed.
The admitted domain was found to be the most extensive for a simple
sinusoidal flow profile with $\tau\simeq 7$
(Fig.~\ref{fig:maps_case1_7}): in this case, practically the whole
upper left half of the $u_0$--$\eta$ plane is allowed. Taken by
themselves, then, the polar field constraints considered in this paper
would suggest the use of this simple flow model with the $\eta/u_0$
ratio exceeding about 50.

{\revision  
It may be worth comparing these results with those of \cite{Whitbread+:SFT}, 
who used the PIKAIA algorithm to optimize the SFT models for different
regimes. The flow profile used in these models differed from our
profiles and was relatively closest to our profile 2. The properties
of the source also differed and the merit criterion was to optimize
overall agreement with observed magnetograms, rather than with the polar field. Their results, 
summarized in their Table~1, indicate admissible ranges for $u_0$ and 
$\eta$ comparable to our results for flow type 2; however, their
preferred ranges for $\tau$ are shorter than ours, in the range 1.5--6
years, except for one configuration of the source term where
significantly higher values were admitted. 
It is to be noted that, as we discussed in Section~3 above, the
neglect of 3D transport processes renders SFT modelling an essentially
phenomenological approach to the description of magnetic flux
transport where the effective flows, diffusivities or decay times
yielding a good fit to observations do not necessarily correspond to
their actual values on the solar surface. In view of this it may come
as little surprise that optimization for different merit functions may
lead to different results. The choice of parameters will simply need
to be adapted to the objectives of the modelling. 
}

A plausible further extension of this work would admit intercycle
variations in cycle amplitudes and periods, focusing on how SFT
parameter choices influence the interplay of stochastic and nonlinear
effects.

\begin{acknowledgements}
This research was supported by the Hungarian National Research,
Development and Innovation Fund (grant no. NKFI K-128384) and by the
European Union's Horizon 2020 research and innovation programme under
grant agreement No. 739500.
\end{acknowledgements}

\bibliographystyle{aa}
\bibliography{SFTparams}

\begin{thebibliography}{44}
\expandafter\ifx\csname natexlab\endcsname\relax\def\natexlab#1{#1}\fi

\bibitem[{{Baumann} {et~al.}(2006){Baumann}, {Schmitt}, \&
  {Sch{\"u}ssler}}]{Baumann+:decayterm}
{Baumann}, I., {Schmitt}, D., \& {Sch{\"u}ssler}, M. 2006, \aap, 446, 307

\bibitem[{Baumann {et~al.}(2004)Baumann, Schmitt, Sch{\"u}ssler, \&
  Solanki}]{baumann2004evolution}
Baumann, I., Schmitt, D., Sch{\"u}ssler, M., \& Solanki, S. 2004, Astronomy \&
  Astrophysics, 426, 1075

\bibitem[{Cameron \& Sch{\"u}ssler(2007)}]{cameron2007solar}
Cameron, R. \& Sch{\"u}ssler, M. 2007, The Astrophysical Journal, 659, 801

\bibitem[{{Cameron} {et~al.}(2010){Cameron}, {Jiang}, {Schmitt}, \&
  {Sch{\"u}ssler}}]{Cameron+:tiltprecursor}
{Cameron}, R.~H., {Jiang}, J., {Schmitt}, D., \& {Sch{\"u}ssler}, M. 2010,
  \apj, 719, 264

\bibitem[{{Chen} \& {Zhao}(2017)}]{Chen+Zhao:highpeak}
{Chen}, R. \& {Zhao}, J. 2017, \apj, 849, 144

\bibitem[{DeVore {et~al.}(1985)DeVore, Sheeley~Jr, Boris, Young~Jr, \&
  Harvey}]{devore1985numerical}
DeVore, C., Sheeley~Jr, N., Boris, J., Young~Jr, T., \& Harvey, K. 1985,
  Australian journal of physics, 38, 999

\bibitem[{Dikpati {et~al.}(2006)Dikpati, De~Toma, \&
  Gilman}]{dikpati2006predicting}
Dikpati, M., De~Toma, G., \& Gilman, P.~A. 2006, Geophysical research letters,
  33

\bibitem[{Hathaway(1996)}]{hathaway1996doppler}
Hathaway, D.~H. 1996, The Astrophysical Journal, 460, 1027

\bibitem[{Hathaway {et~al.}(1994)Hathaway, Wilson, \&
  Reichmann}]{hathaway1994shape}
Hathaway, D.~H., Wilson, R.~M., \& Reichmann, E.~J. 1994, Solar Physics, 151,
  177

\bibitem[{Iida(2016)}]{iida2016tracking}
Iida, Y. 2016, Journal of Space Weather and Space Climate, 6, A27

\bibitem[{{Iijima} {et~al.}(2017){Iijima}, {Hotta}, {Imada}, {Kusano}, \&
  {Shiota}}]{Iijima+:plateau}
{Iijima}, H., {Hotta}, H., {Imada}, S., {Kusano}, K., \& {Shiota}, D. 2017,
  \aap, 607, L2

\bibitem[{{Imada} \& {Fujiyama}(2018)}]{Imada+:midpeak}
{Imada}, S. \& {Fujiyama}, M. 2018, \apjl, 864, L5

\bibitem[{{Jiang} {et~al.}(2011){Jiang}, {Cameron}, {Schmitt}, \&
  {Sch{\"u}ssler}}]{Jiang+:1700a}
{Jiang}, J., {Cameron}, R.~H., {Schmitt}, D., \& {Sch{\"u}ssler}, M. 2011,
  \aap, 528, A82

\bibitem[{{Jiang} \& {Cao}(2018)}]{Jiang+Cao}
{Jiang}, J. \& {Cao}, J. 2018, Journal of Atmospheric and Solar-Terrestrial
  Physics, 176, 34

\bibitem[{Jiang {et~al.}(2014)Jiang, Hathaway, Cameron, Solanki, Gizon, \&
  Upton}]{jiang2014magnetic}
Jiang, J., Hathaway, D., Cameron, R., {et~al.} 2014, Space Science Reviews,
  186, 491

\bibitem[{Jiang {et~al.}(2018)Jiang, Wang, Jiao, \&
  Cao}]{jiang2018predictability}
Jiang, J., Wang, J.-X., Jiao, Q.-R., \& Cao, J.-B. 2018, The Astrophysical
  Journal, 863, 159

\bibitem[{{Lemerle} \& {Charbonneau}(2017)}]{Lemerle2}
{Lemerle}, A. \& {Charbonneau}, P. 2017, \apj, 834, 133

\bibitem[{Lemerle {et~al.}(2015)Lemerle, Charbonneau, \&
  Carignan-Dugas}]{Lemerle1}
Lemerle, A., Charbonneau, P., \& Carignan-Dugas, A. 2015, The Astrophysical
  Journal, 810, 78

\bibitem[{{Lin} \& {Chou}(2018)}]{Lin+Chou:cycledep.flows}
{Lin}, C.-H. \& {Chou}, D.-Y. 2018, \apj, 860, 48

\bibitem[{{Orozco Su{\'a}rez} {et~al.}(2007){Orozco Su{\'a}rez}, {Bellot
  Rubio}, {del Toro Iniesta}, {Tsuneta}, {Lites}, {Ichimoto}, {Katsukawa},
  {Nagata}, {Shimizu}, {Shine}, {Suematsu}, {Tarbell}, \& {Title}}]{Orozco2007}
{Orozco Su{\'a}rez}, D., {Bellot Rubio}, L.~R., {del Toro Iniesta}, J.~C.,
  {et~al.} 2007, \apjl, 670, L61

\bibitem[{Pesnell(2008)}]{pesnell2008predictions}
Pesnell, W.~D. 2008, Solar Physics, 252, 209

\bibitem[{{Petrie}(2015)}]{Petrie:LRSP}
{Petrie}, G.~J.~D. 2015, Living Reviews in Solar Physics, 12, 5

\bibitem[{{Petrovay}(2010)}]{Petrovay:LRSP}
{Petrovay}, K. 2010, Living Reviews in Solar Physics, 7, 6

\bibitem[{{Petrovay}(2019)}]{Petrovay:LRSP2}
{Petrovay}, K. 2019, arXiv e-prints, arXiv:1907.02107

\bibitem[{Petrovay \& Szak\'aly(1999)}]{Petrovay+Szakaly:2d.pol}
Petrovay, K. \& Szak\'aly, G. 1999, \solphys, 185, 1

\bibitem[{Schad {et~al.}(2011)Schad, Timmer, \& Roth}]{schad2011unified}
Schad, A., Timmer, J., \& Roth, M. 2011, The Astrophysical Journal, 734, 97

\bibitem[{{Schrijver} {et~al.}(2002){Schrijver}, {De Rosa}, \&
  {Title}}]{schrijver2002missing}
{Schrijver}, C.~J., {De Rosa}, M.~L., \& {Title}, A.~M. 2002, \apj, 577, 1006

\bibitem[{{Schrijver} \& {Zwaan}(2000)}]{Schrijver+Zwaan:book}
{Schrijver}, C.~J. \& {Zwaan}, C. 2000, {Solar and Stellar Magnetic Activity}
  (Cambridge Univ. Press)

\bibitem[{Sheeley {et~al.}(1983)Sheeley, Boris, Young, DeVore, \&
  Harvey}]{sheeley1983quantitative}
Sheeley, N., Boris, J., Young, T., DeVore, C., \& Harvey, K. 1983, in
  Symposium-International Astronomical Union, Vol. 102, Cambridge University
  Press, 273--278

\bibitem[{Sheeley {et~al.}(1989)Sheeley, Wang, \&
  DeVore}]{sheeley1989implications}
Sheeley, N., Wang, Y.-M., \& DeVore, C. 1989, Solar physics, 124, 1

\bibitem[{{Sheeley}(2005)}]{Sheeley:LRSP}
{Sheeley}, Jr., N.~R. 2005, Living Reviews in Solar Physics, 2, 5

\bibitem[{{Sheeley} {et~al.}(1989){Sheeley}, {Wang}, \&
  {DeVore}}]{Sheeley+:topknot}
{Sheeley}, Jr., N.~R., {Wang}, Y.-M., \& {DeVore}, C.~R. 1989, \solphys, 124, 1

\bibitem[{{Stix}(2004)}]{Stix:book}
{Stix}, M. 2004, {The sun : an introduction} ({Springer})

\bibitem[{Svalgaard {et~al.}(1978)Svalgaard, {Duvall}, \&
  {Scherrer}}]{Svalgaard+:polarfield}
Svalgaard, L., {Duvall}, Jr., T., \& {Scherrer}, P. 1978, Solar Phys., 58, 225

\bibitem[{van Ballegooijen {et~al.}(1998)van Ballegooijen, Cartledge, \&
  Priest}]{van1998magnetic}
van Ballegooijen, A., Cartledge, N., \& Priest, E. 1998, The Astrophysical
  Journal, 501, 866

\bibitem[{{Virtanen} {et~al.}(2017){Virtanen}, {Virtanen}, {Pevtsov}, {Yeates},
  \& {Mursula}}]{Virtanen+:SFT}
{Virtanen}, I.~O.~I., {Virtanen}, I.~I., {Pevtsov}, A.~A., {Yeates}, A., \&
  {Mursula}, K. 2017, \aap, 604, A8

\bibitem[{Wang {et~al.}(2000)Wang, Lean, \& Sheeley~Jr}]{wang2000long}
Wang, Y.-M., Lean, J., \& Sheeley~Jr, N. 2000, Geophysical Research Letters,
  27, 505

\bibitem[{Wang {et~al.}(1989)Wang, Nash, \& Sheeley~Jr}]{wang1989evolution}
Wang, Y.-M., Nash, A., \& Sheeley~Jr, N. 1989, The Astrophysical Journal, 347,
  529

\bibitem[{{Wang} \& {Sheeley}(1994)}]{Wang+:patternrot}
{Wang}, Y.~M. \& {Sheeley}, N.~R., J. 1994, \apj, 430, 399

\bibitem[{{Wang} {et~al.}(2002){Wang}, {Sheeley}, \& {Lean}}]{Wang:flowvar}
{Wang}, Y.-M., {Sheeley}, Jr., N.~R., \& {Lean}, J. 2002, \apj, 580, 1188

\bibitem[{{Whitbread} {et~al.}(2019){Whitbread}, {Yeates}, \&
  {Mu{\~n}oz-Jaramillo}}]{Whitbread+:disconn}
{Whitbread}, T., {Yeates}, A.~R., \& {Mu{\~n}oz-Jaramillo}, A. 2019, \aap, 627,
  A168

\bibitem[{{Whitbread} {et~al.}(2017){Whitbread}, {Yeates},
  {Mu{\~n}oz-Jaramillo}, \& {Petrie}}]{Whitbread+:SFT}
{Whitbread}, T., {Yeates}, A.~R., {Mu{\~n}oz-Jaramillo}, A., \& {Petrie},
  G.~J.~D. 2017, \aap, 607, A76

\bibitem[{{Wilcox} {et~al.}(1970){Wilcox}, {Schatten}, {Tanenbaum}, \&
  {Howard}}]{Wilcox+:patternrot}
{Wilcox}, J.~M., {Schatten}, K.~H., {Tanenbaum}, A.~S., \& {Howard}, R. 1970,
  \solphys, 14, 255

\bibitem[{{Zhao} {et~al.}(2014){Zhao}, {Kosovichev}, \&
  {Bogart}}]{Zhao+:lowpeak}
{Zhao}, J., {Kosovichev}, A.~G., \& {Bogart}, R.~S. 2014, \apj, 789, L7

\end{thebibliography}

\end{document}